\documentclass[article, 11pt,]{amsart}
\pdfoutput=1
\usepackage[english]{babel}
\usepackage{graphicx,caption}
\usepackage[utf8]{inputenc}
\usepackage[T1]{fontenc}
\usepackage{float,hyphenat}
\usepackage[top=2.5cm,bottom=2.5cm,left=2.5cm,right=2.5cm,marginparwidth=2.5cm]{geometry}
\usepackage{amsmath}
\usepackage{color}
\usepackage[version=4]{mhchem} 
\usepackage[colorinlistoftodos]{todonotes}
\usepackage{xcolor}
\usepackage{siunitx}
\usepackage[section]{placeins} 
\usepackage[backend=biber,style=numeric-comp,sorting=none]{biblatex}
\usepackage{setspace}
\title{Selective Undercut of Undoped Optical Membranes for Spin-Active Color Centers in 4H-SiC}

\bibliography{references.bib}
\begin{document}
    \maketitle
    \begin{center}    
       Jonathan Dietz, Amberly Xie, Aaron M. Day, Evelyn L. Hu\\
      \small John A. Paulson School of Engineering and Applied Sciences\\ Harvard University, Cambridge, MA. 02138, USA 
    \end{center}
    \begin{abstract}
      Silicon carbide (SiC) is a semiconductor used in quantum information processing, microelectromechanical systems, photonics, power electronics, and harsh environment sensors. However, its high temperature stability, high breakdown voltage, wide bandgap, and high mechanical strength are accompanied by a chemical inertness which makes complex micromachining difficult. Photoelectrochemical etching is a simple, rapid means of wet processing SiC, including the use of dopant selective etch stops that take advantage of mature SiC homoepitaxy. However, dopant selective photoelectrochemical etching typically relies on highly doped material, which poses challenges for device applications such as quantum defects and photonics that benefit from low doping to produce robust emitter properties and high optical transparency. In this work, we develop a new, selective photoelectrochemical etching process that relies not on high doping but on the electrical depletion of a fabricated diode structure, allowing the selective etching of an n-doped substrate wafer versus an undoped epitaxial ($N_a=1(10)^{14}cm^{-3}$) device layer. We characterize the photo-response and photoelectrochemical etching behavior of the diode under bias and use those insights to suspend large ($>100\mu m^2$) undoped membranes of SiC. We further characterize the compatibility of membranes with quantum emitters, performing comparative spin spectroscopy between undoped and highly doped membrane structures, finding the use of undoped material improves ensemble spin lifetime by $>3x$. This work enables the fabrication of high-purity suspended thin films suitable for scalable photonics, mechanics, and quantum technologies in SiC.
    \end{abstract}

\section{Introduction}
A high-yield, systematic means of producing high-quality thin-films is an outstanding challenge in realizing scalable quantum and classical technology. The availability of single crystal thin films with high purity and optical quality has led to large advances in miniaturizing typically bulk optical components with equivalent integrated photonic devices and the development of micro-electromechanical circuits.\cite{qu_cmos_2016, zheng_high-quality_2019, zhu_integrated_2021,,guo_tunable_2021} One such approach to creating thin films in semiconductor materials with effective p and n-type doping, is dopant selective photoelectrochemical (PEC) or electrochemical etching of one doping type over the other in an epitaxial or implanted homojunction.\cite{linden_fabrication_1989} Such etching strategies have been designed in silicon, gallium nitride, gallium arsenide, and silicon carbide homojunctions.\cite{eijkel_new_1990,shor_dopant-selective_1997, youtsey_dopant-selective_1998,khare_dopant_1991} However, these approaches rely on heavily doped device layers, as lowering the doping of the device layer reduces etching selectivity.\cite{khare_dopant_1991} A dopant selective electrochemical etch that does not require high device layer doping is desirable in applications where high doping degrades performance or design flexibility, such as electronics, photonics, and mechanics or in material systems where effective doping is not available for either p-type or n-type, such as zinc oxide or diamond respectively.\cite{soref_electrooptical_1987,mihailovich_low_1992,janotti_fundamentals_2009,yang_progress_2022,}

This work presents such a process, with a particular application to Silicon Carbide (SiC), a wide band-gap semiconductor widely studied for use in photonics, mechanics, and quantum information\cite{lukin_integrated_2020,long_4h-silicon_2023, zhang_material_2020, dietz_spin-acoustic_2023}. Consequently, SiC is such a material that could benefit from a PEC etching process that results in a low doped thin film. For example, SiC hosts several solid-state quantum defect qubits which show lifetime limited optical linewidths, long spin coherence, and high initialization fidelity which could form the basis of quantum memory nodes in a quantum network.\cite{nagy_high-fidelity_2019, anderson_electrical_2019} Dopant-selective PEC has successfully created suspended photonics in SiC with integrated defects, yet these devices and defects demonstrate degraded optical and spin performance, respectively due to the required use of high doping concentration.\cite{magyar_high_2014,bracher_selective_2017,crook_purcell_2020} This requirement is predicated on the ability to independently control band bending in each material doping type with respect to the chemical potential of decomposition in the etching electrolyte, which is more favorable in layers with large differences in doping.\cite{pavunny_doping_2019} 

We present a modified PEC process, that uses the bias of a simple transparent Shottky barrier diode to enable independent electrochemical control of each material layer. We show selective etching between an undoped layer of 4H-SiC and a highly doped substrate, demonstrating high etch rates ($>4\mu m/hr$), low surface roughness ($\sim1nm$ RMS), and large areas ($>100\mu m^2$) of undercut material. We further demonstrate the immediate device performance advantages of this process over a process that relies high doping of the device layer alone, comparing the quantum spin performance of ensembles of silicon monovacancies in doped and undoped layers . When an undoped layer is used, lower strain, longer spin coherence, and lower inhomogeneous broadening is observed. The reliance of this technique on an externally fabricated diode device rather than solely on material properties indicates that this technique could be adapted to other material systems where junctions can be fabricated.

\section{Main}
\begin{figure*}[t]
    \includegraphics[scale = 0.98]{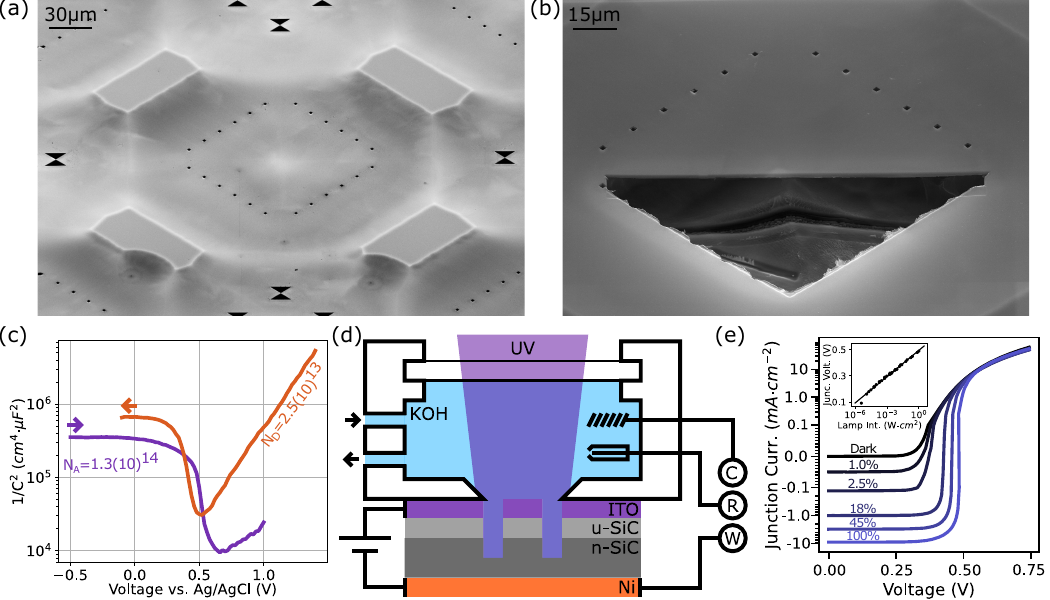}
        \caption{(a) Scanning electron microscope (SEM) image of a $100\mu m^2$ membrane with support structures imaged at 30keV to reveal substrate structure. (b) focused ion beam (FIB) cross section of membrane showing uniform membrane thickness. (c) Electrochemical impedance spectroscopy of an undoped, $500nm$ thick epilayer of SiC showing low p-type conductivity, determined from linear fitting ($1/C^2$) and measured using an Al contact. (d) configuration of the PEC cell used to perform CV and etching of 4H-SiC structures. $C$, $R$, and $W$ are the Pt counter, Ag/AgCl reference, and SiC working electrodes respectively. u-SiC denotes the undoped nature of the p$\textsuperscript{-}$ layer (e) I-V curves of the photodiode formed by the ITO contacted undoped layer and Ni contacted substrate. Inset: illumination dependent photo-biasing of the diode.
        }
    \label{F1}
\end{figure*}
Due to its highly chemically inert nature, suspended SiC photonic and mechanical devices are often created using SmartCut, grind-and-polish, deep backside etching, or via dopant selective photo-electro-chemical etching (PEC).\cite{zheng_high-quality_2019, bracher_fabrication_2015, jiang_semi-insulating_2021, lukin_4h-silicon-carbide--insulator_2020, heiler_spectral_2024} These techniques have enabled fabrication of high-quality devices, but performance is typically limited by wafer non-uniformity, yield, excess strain and charge noise. PEC usually relies on high doping contrast between epitaxial layers to selectively etch, for example, a highly n-doped substrate from a comparably p-doped epilayer, with the epilayer as the ultimate host of both photonic devices and quantum defects. However, free carrier absorption from highly doped material and charge noise creates optical propagation loss and limits qubit spin coherence.\cite{crook_purcell_2020} Previously fabricated devices demonstrated high quality optical resonances, but suffered from high strain, roughened surfaces and poor defect coherence.\cite{crook_purcell_2020, dietz_optical_2022} 

The photoelectrochemical etching of single crystal SiC is well understood for both n and p doping types in alkaline solutions. In both cases, etching occurs through the elimination of silicon through the formation of a soluble silicate product and carbon monoxide via the charge exchange of holes with hydroxide ions in solution\cite{dorp_anodic_2007}:
\begin{equation}
SiC+8OH^- + 6h^+ \rightarrow [Si(OH)_2O_2]^{2-} +CO + 3H_2O
\end{equation}

In p-type material, electrochemical etching can occur without illumination, because holes are a majority carrier in the material.\cite{adachi_single-crystalline_2013} Because the valence band in p-type material typically bends in energy downwards towards the chemical potential of the electrolyte, the onset of etching occurs when the cell bias lowers the barrier holes experience in reaching this interface.  By contrast, in n-type material, etching does not occur unless the SiC is illuminated to generate the requisite holes, and though the onset of etching can occur without cell bias, it is often beneficially controlled by the application of such a bias.\cite{dorp_anodic_2007} Etching rate is limited by the rate of reaction at the semiconductor-electrolyte interface, controlled by the quantity and diffusion rate of ions in solution and holes in the semiconductor towards the interface. The diffusion rate of ions to the interface is controlled by mass transport in the bulk of the solution and the potential drop that ions experience at the Helmholtz double layer. In the semiconductor, the diffusion of holes to the interface is controlled by the built-in voltage of the semiconductor-electrolyte junction and the voltage difference between the semiconductor (working) electrode and the counter electrode. The number of holes is largely determined by the intensity of illumination. Photoelectrochemical etching of silicon carbide is stable if anodic oxidation does not outpace the dissolution by charge exchange, in which case the etch is passivated by an insulating layer of $SiO_2$.\cite{dorp_anodic_2007} Additionally, the etching rate can be controlled by crystallography. In hexagonal polytypes of SiC, the carbon-face of the material is known to be significantly more reactive than the silicon-face of the material and consequently has a significantly faster PEC etch.\cite{kato_electrochemical_2003} p-type (n-type) material may be selectively etched against relatively n-type (p-type) material by appropriate selection of cell bias by setting a more negative (positive) cell bias which fixes the energy position of the valence band relative to the electrolyte`s chemical potential, potentially yielding suspended structures if the substrate is targetted in the etch (Figure \ref{F1}a-b). In a homojunction that combines both doping types this relies on a high doping difference (e.g. $N_a=1(10)^{14}cm^{-3}$ vs. $N_d=1(10)^{14}cm^{-3}$) and allows a $0.5V$ potential window for selectively etching relatively thick layers of SiC.\cite{pavunny_doping_2019}

\begin{figure*}[t]
    \centering
    \includegraphics[scale = 1]{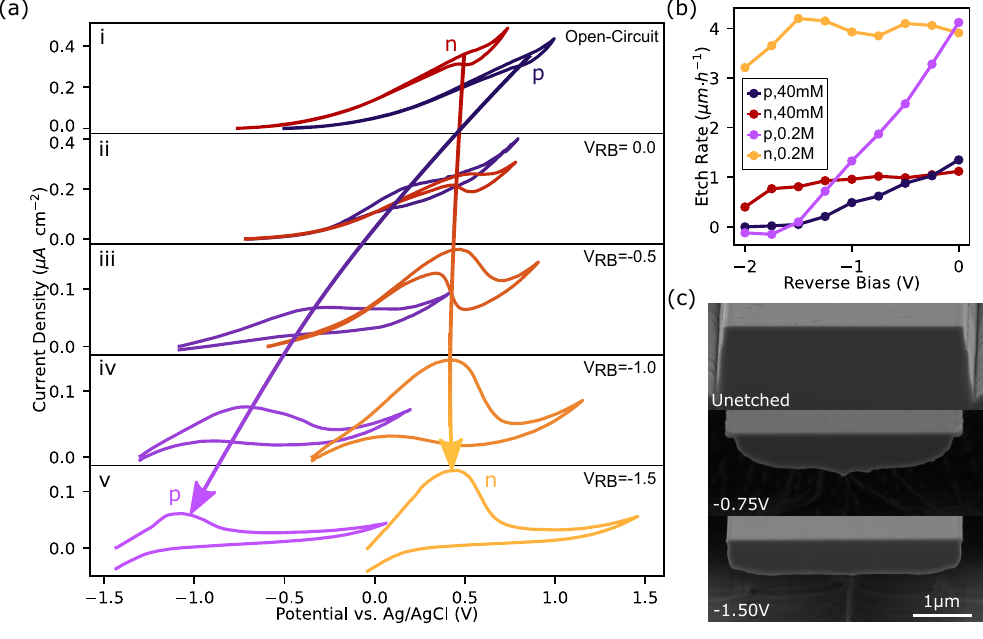}
        \caption{(a) Cyclic voltamograms ($20mVs^{-1}$) of etched mesa structures under open circuit and various diode reverse biases, measured on the n-type contact (red, orange, and yellow) or p-type contact (indigo, purple). (b) Etch rates for different KOH concentrations as a function of diode reverse bias. (c) FIB cross section of mesa structures before PEC etching and after 2 hours of PEC etching with a working electrode bias of 0.2V and an partially depleted ($-0.75V$) and fully depleted ($-1.5V$) reverse bias.
        }
    \label{F2}
\end{figure*}

Unfortunately, in addition to the constraints on applications imposed by the high doping spread necessary for selective etching, doping selectivity in a homojunction depends on the ability to mutually bias both doping types in the cell. While this assumption is reasonable in thick layers or under low illumination conditions, it breaks down in epilayer thicknesses approaching useful thicknesses in photonic devices because of strong photo-biasing of the homojunction. To assess this effect, we construct a junction from a thin, $500nm$ undoped layer of 4H-SiC epitaxially deposited onto an n-type ($3(10)^{18}$ nitrogen doped) substrate wafer. Contact is made to the n-type substrate by thermally evaporating $200nm$ of nickel on the substrate backside followed by subsequent rapid thermal annealing for 5 minutes at 950$^{\circ}C$ in an argon atmosphere. Contact is made to the unintentional doped layer by thermally evaporating $200nm$ of aluminum onto the epilayer, with a circular aperture in the center of the sample left masked. The mask is then stripped, and contact is made by rapid thermal annealing for 5 minutes at 950$^{\circ}C$ in an argon atmosphere. The sample is then immersed in $2M$ KCl solution with the Al contact covered by a silicone o-ring. The carrier concentration and conductivity type of the epilayer is determined using electrochemical impedance spectroscopy-based Mott Schottky analysis\cite{sivula_mottschottky_2021}:

\begin{equation}
\frac{1}{C^{2}_{SC}}=\frac{2}{\epsilon_0 \epsilon_r N_{a,d} e}(V-V_{FB}-\frac{k_B T}{e})
\end{equation}

\noindent where $C_{SC}$ is the capacitance of the space charge region, $ \epsilon_{0}$ and $ \epsilon_{r}$ are the vacuum and relative (9.6 for 4H-SiC) permittivities, $N_{a,d}$ is the acceptor or donor concentration, respectively. $e$ is the electron charge, $k_B$ is the Boltzmann constant and $T$ is the temperature in degrees Kelvin. By linear fitting to the slope of the measured capacitance in cathodic and anodic sweeps, it is possible to determine the acceptor and donor concentrations, respectively. The net doping is found to be $1.15(10)^{14}cm^{-3}$, as shown in Figure \ref{F1}c, hereafter referred to as the "p-type" layer. 

For PEC-enabled release of etched structures, indium tin oxide (ITO) is adopted as an optically transparent contact to the p-type layer through which the underlying SiC can be illuminated while still providing uniform electrical connectivity across the cell. For diode characterization, cyclic voltammetry, and PEC etching, ITO is sputtered at a rate of $3nm/min$ to form a $100nm$ layer, ramping the substrate to 600$^{\circ}C$ over the course of deposition. Substrate heating was found to be necessary to form a suitably conductive contact to the p-type epilayer. Illumination for all measurements and etching is provided by a $500W$ Xe-Hg Arc lamp with its visible spectrum filtered by a $400nm$ short-pass filter to reduce excess heating of the sample, producing a peak intensity of $1W\dot cm^{-2}$. Before fabricating membrane devices we first characterize the photo-response of the diode structure, to determine the degree to which the epilayer is photo-biased as the lamp intensity is varied. In the dark, the current-voltage (IV) curve shows a turn-on voltage of $0.5V$. Under increasing illumination, the diode shows a typical photodiode response, with a peak photo-bias of $490mV$ measured at a lamp intensity of $1W\dot cm^{-2}$ (Figure \ref{F1}e). 

To characterize the dopant-selective undercut process, mesas are first defined by patterning the ITO layer using positive tone (PMMA495 C6) electron-beam lithography. The pattern is transferred to the ITO by inductively coupled plasma reactive ion etching in H2/CF4, and then subsequently transferred into the SiC to a depth of $2\mu m$ using dry etching in SF6/O2. ITO is found to be a highly selective hard mask, as oxide reduction is continually countered by plasma oxidation via the oxygen present in the SiC etching gas mixture, yielding a selectivity of 50:1. PEC etching is performed in $40-500mM$ potassium hydroxide (KOH) electrolyte which is continuously circulated from a 10$^{\circ}C$ reservoir to ensure a constant electrolyte concentration and to extract waste heat, with no supporting electrolyte added. The counter electrode is a wound Pt wire and the reference electrode is an Ag/AgCl electrode in 3M KCl. The membranes are undercut using a PEC cell, with two separate electrical circuits present. The first is the potentiostat, used to bias either the n-type or p-type SiC as a working electrode against the Pt counter electrode. The second is the junction bias, controlled by attaching a voltage source to the p-type and n-type contacts. To assess the impact of photo-biasing on the etching kinetics of the junction, cyclic voltammetry (CV) is performed under various junction bias conditions with a sweep rate of $20mVs^{-1}$ (Figure \ref{F2}a). Under full-illumination ($1W\dot cm^{-2}$) and no external bias applied to the junction, the p-type and n-type CV curves recorded are similar except that the p-type curve is shifted anodic relative to the n-type curve (Figure \ref{F2}a.i). This is contrary to rigorous studies of thick layers of 4H-SiC which found that n-type curves should be anodic relative to p-type curves.\cite{pavunny_doping_2019} Furthermore, no distinct oxidation or reduction peaks are observed corresponding to p-type or n-type etching, confirming that the strong photo-biasing of the cell has eliminated etch selectivity. However, by noting that the photo-biasing occurs because of the presence of the junction formed by the epilayer and substrate, it can be concluded that the photo-biasing can be completely suppressed by reverse biasing the junction to correct for the photo-generated voltage. By closing the junction circuit between the ITO and Ni junction contacts, the CV curves converge (Figure \ref{F2}a.ii). While closing the junction circuit may restore selectivity in highly doped material, we must further increase the junction bias in our case of an undoped epilayer. The junction bias depletes the epilayer of carriers and creates selectivity between regions of different doping where it is not possible to achieve rapid selective etching without diode biasing. As greater reverse bias is progressively applied, the quasi-Fermi level in the valence band of the p-type layer rises relative to that in the n-type layer and the dissolution potential into the electrolyte, and the p-type CV curve shifts cathodic (Figure \ref{F2}a.iii-v). At the same time, the characteristic oxidation and reduction peaks associated with controlled etching reappear, now at cell biases significantly different enough to support selective etching. 
\begin{figure}[b]
    \includegraphics[scale = 1]{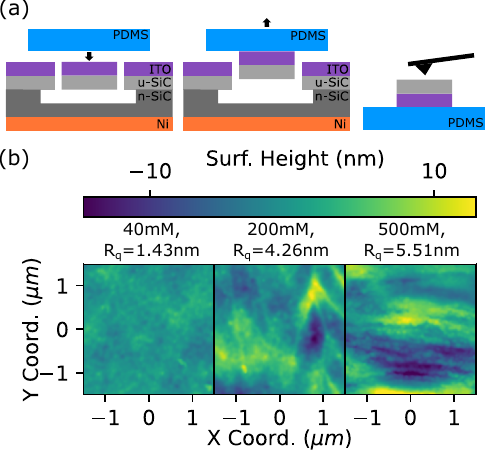}
        \caption{(a) PDMS membrane release process for AFM characterization (b) AFM height maps for different electrolyte concentrations with corresponding root mean squared roughness values ($R_q$) after a 2 hour release process with $0.2V$ cell bias and $1W\dot cm^2$ illumination.
        }
    \label{F3}
\end{figure}

\begin{figure*}[t]
    \includegraphics[scale = 1]{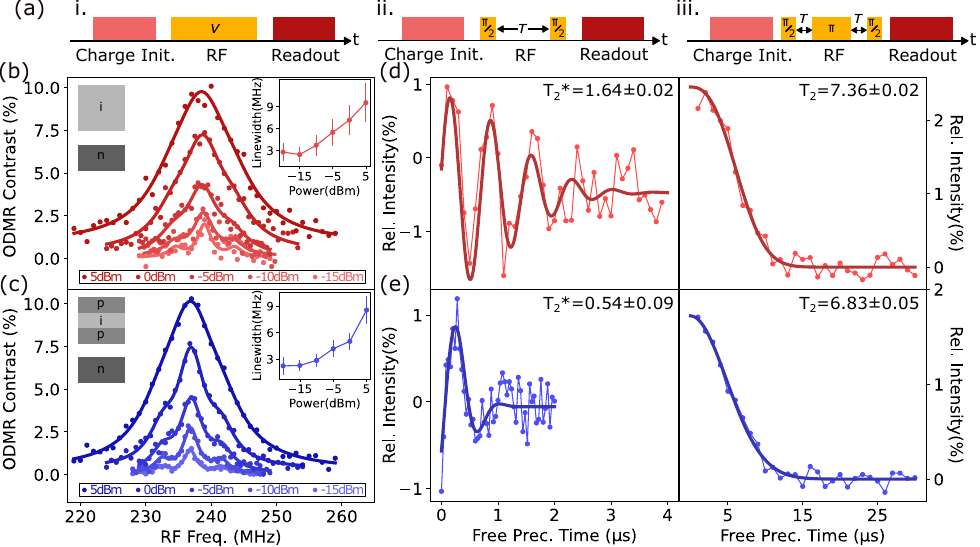}
        \caption{Spin spectroscopy of weak ensembles of $V_{Si}^-$ in (red) undoped epilayer and (blue) p-i-p doped structures. (a) Pulse sequences used for (i.) ODMR, (ii.) FID and (iii.) spin dephasing measurements. (b-c) ODMR spectrums with fitted triple Lorentzian spectrum based on natural $^{29}Si$ concentration in the nearest neighbouring site of the defects. Inset: Fitted power dependent line-width of the central peak. (d-e) Free induction decay ($T_2^{*}$) fitted to a decaying sinusoid with an envelope of $exp((-t/T_2^{*})^2)$ measured via a Ramsey sequence $2MHz$ off-resonance and spin dephasing time ($T_2$) fitted to an exponential decay of $exp((-t/T_2)^2)$ by Hahn echo sequence on-resonance. All measurements performed at $6K$.
        }
    \label{F4}
\end{figure*}

To assess doping selectivity in realistic structures, a series of $4\mu m$ wide mesas are undercut in $40mM$ and $200mM$ KOH at different junction reverse biases (Figure \ref{F2}), with a cell bias of $0.2V$, selected to ensure that the n-type electrode is not passivated.\cite{dorp_anodic_2007} Lateral etching rates and membrane morphology are determined using focused-ion-beam (FIB) cross sectioning. In both electrolyte concentrations, as the reverse bias increases, p-type etching is suppressed before becoming fully passivated, past a reverse bias of $-1.3V$(Figure \ref{F2}b). Cross sectional images corroborate this finding, showing partial undercutting of the p-type layer at moderate reverse biases ($-0.75V$), and full passivation at high reverse biases($-1.5V$) (Figure \ref{F2}c). This result demonstrates that at sufficient reverse bias, diffusion of photo-generated carriers can be predominantly controlled not by differences in doping but by external biasing. 

Smoothly-etched surfaces are highly desirable for photonic and mechanical structures, as scattering loss and surface damping are therefore reduced. We investigate the impact that electrolyte concentration has on the surface roughness of the underside of our membranes. The membranes shown in Figure \ref{F2}c are fully suspended and then lifted out using polydimethylsiloxane (PDMS) stamping (Figure \ref{F3}a). The surface roughness of the membrane underside is then measured by tapping mode atomic force microscopy. We find that the concentration of electrolyte strongly controls the surface quality of the underside of the membranes, with lower concentration ($40mM$ KOH) yielding a RMS surface roughness value of $1.43nm$ (Figure \ref{F3}b-d), whereas higher concentrations of KOH yield rougher surfaces ($5.51nm$ for $500mM$ KOH). Previous work suggests that roughness increases in a regime where oxide formation exceeds the rate of oxide removal, indicating this balance may be more difficult to establish at high electrolyte concentrations.\cite{pavunny_doping_2019} Similar work also showed that above $500mM$ concentrations, only porous SiC is formed by PEC.\cite{kato_electrochemical_2003}

To assess the potential for the material to host quantum spin registers, the reverse bias undercut process is paired with spin resonance and spin lifetime measurements. A suspended,  500nm thick undoped membrane formed using the process described, etched in $40mM$ KOH, is compared to another,\\ $500 nm$ thick suspended membrane comprised of  p-i-p-n ($150nm$ p-type, $200nm$ undoped, $150nm$ p-type, n-type substrate) material used in previous work, that is undercut under identical conditions.\cite{dietz_optical_2022} Prior work found that slight shifts in the ZPL, ODMR and spin coherence times in photoelectrochemical etched structures was a result of high doping, porosity and strain induced by the PEC process.\cite{crook_purcell_2020} We select the silicon monovacancy for study as it can be flexibly introduced and has been shown to have high quality optical and spin properties in nanostructures.\cite{day_laser_2023, babin_fabrication_2022} Each sample is identically implanted with He ions of energy $45keV$ at a dose of $10^{11}\,ions\cdot cm^{-2}$  followed by quenching from 650$^{\circ}C$ after annealing in air for 30 minutes, to form a weak ensemble of silicon monovacancies ($V_{Si}^-$). The samples are then loaded into a $4K$ flow cryostat and fitted with a microwave wire antenna to excite ground state spin resonances. The charge state of the defects is initialized by an off resonant helium neon laser with a power of $3mW$, probed by $28dbm$ of input RF power, and photoluminescence is measured on the phonon sideband (>$925nm$)  of the defects using near resonant excitation with a $916nm$ external cavity diode laser with a power of $100\mu W$.(Figure \ref{F4}a)

To compare the spin performance of emitters in both the heavily doped and undoped material, power dependent optically detected magnetic resonance \\(ODMR), free induction decay times ($T_2^{*}$), and spin dephasing times ($T_2$) are measured for each ensemble (Figure \ref{F4}a). The ground state ($S=3/2$) spin levels, which are typically separated by a zero-field splitting of $70MHz$ are separated further by a moderate magnetic field of $60G$ to a final splitting of $240MHz$ to reduce the effect of heteronuclear spin flip processes that limit the $T_2$ and $T_2^{*}$ times of ensemble of emitters. Assessing the $m_s=-\frac{1}{2}$ to $m_s=-\frac{3}{2}$ transition, ODMR studies in Figure \ref{F4}b,c show similar performance, with emitters in both material types demonstrating narrow inhomogeneous broadening that saturate to a minimum at similar drive powers. The $T_2^{*}$ time of $1.64\pm0.02\mu s$ measured in the undoped material (Figure \ref{F4}d) is longer than that in the doped material (Figure \ref{F4}e), approaching the theoretically predicted low field limit of ($1.9\mu s$) for monovacancy spins in a naturally abundant $^{29}Si$,$^{13}C$ nuclear bath.\cite{yang_electron_2014} This demonstrates that the use of undoped material reduces the presence of influences that induce fast dephasing of the spins, which previously limited PEC-etched device performance \cite{crook_purcell_2020}. In contrast, the $T_2$ time is similar between the two samples, showing that in both cases on longer timescales the defects are limited by similar slow, dephasing processes. This implies that doping or material defects, and not material porosity or fabrication damage, primarily drives the degradation of $V_{Si}^-$ spin properties in PEC processed material.

\section{Conclusion}
We have demonstrated a means of dopant-selective PEC etching that significantly relaxes constraints placed on doping density to achieve etch selectivity--thereby enabling the suspension of low-surface roughness, large-area membranes with an improved charge-noise environment to host quantum emitters. This approach reduces the epitaxial steps necessary for high purity, PEC-suspended photonics and mechanics in SiC to one step with the addition of a commonly used hard-mask (ITO) and a circuit to bias the junction formed between the epilayer and substrate. We investigate the photo-electrochemical attributes of SiC diodes under external biasing and show that smooth membranes of undoped silicon carbide can be suspended at rates exceeding $4um/hr$. We additionally note that the demonstrated 500nm device layer is defined by epitaxial growth and can therefore be flexibly changed to suit application. Finally, we perform comparative spin spectroscopy on suspended doped and undoped materials, demonstrating that undoped material beneficially reduces the effect of fast spin noise on quantum defects and is compatible with PEC processing. The development of a means of reliably suspending large undoped membranes of SiC opens up new opportunities leveraging 4H-SiC's high yield strength for nanomechanical oscillators and applications that integrate quantum emitters into electronic and photonic devices.\cite{sementilli_ultralow_2024} Beyond SiC, this technique relaxes the material constraints for dopant selective etching, suggesting a means of selectively etching other inert material where only a single doping type can be reliably created, such as diamond.

\section{Methods}

    PEC etching is performed in $40-500mM$ potassium hydroxide (KOH) electrolyte which is continuously circulated from a 10$^{\circ}C$ reservoir to ensure a constant electrolyte concentration and to extract waste heat, with no supporting electrolyte added. The counter electrode is a wound Pt wire and the reference electrode is an Ag/AgCl electrode in 3M KCl. For spin measurements, each sample is identically implanted with He ions of energy $37keV$ at a dose of $10^{11}\,ions\cdot cm^{-2}$  followed by quenching from 650$^{\circ}C$ after annealing in air for 30 minutes, to form a weak ensemble of silicon monovacancies ($V_{Si}^-$). The samples are then loaded into a $4K$ flow cryostat and fitted with a microwave wire antenna to excite ground state spin resonances. The charge state of the defects is initialized by an off resonant helium neon laser with a power of $3mW$, and photoluminescence is measured on the phonon sideband (>$925nm$)  of the defects using near resonant excitation with a $916nm$ external cavity diode laser with a power of $100\mu W$. 

\section{Acknowledgements}
 Research reported in this publication was supported as part of the AWS Center for Quantum Networking’s research alliance with the Harvard Quantum Initiative (or HQI). Portions of this work were performed at the Harvard University Center for Nanoscale Systems (CNS); a member of the National Nanotechnology Coordinated Infrastructure Network (NNCI), which is supported by the National Science Foundation under NSF award no. ECCS-2025158. A.M.D. acknowledges funding from the Science and Technology Center for Integrated Quantum Materials, NSF Grant No. DMR-1231319. J.R.D acknowledges funding from NSF RAISE-TAQS Award 1839164. 

\onecolumn
\printbibliography

\end{document}